\documentclass[twocolumn]{article}    

\title{WORLD-SHEET OF THE DISCRETE LIGHT FRONT STRING}  
\author{Gordon W. Semenoff \footnote{On leave from Department of Physics and Astronomy, University of British Columbia, Vancouver.}
\\~~\\ Niels Bohr Institute, Blegdamsvej 17, DK2100
Copenhagen \O, Denmark\\email: semenoff@nbi.dk\\~~\\Proceedings of ICHEP2000, Osaka}

\begin{document}           

\maketitle                 

\abstract{ Some aspects of light-like compactifications
of superstring theory and their implications
for the matrix model of M-theory are discussed.}








\section{Light-Like compactification}\label{sec:light}

T-duality is one of the most profound of stringy phenomena.  Closed
string theory on a space-time which has a compact dimension with
radius $R$ has the same spectrum as a string theory on a space with
radius $
\alpha'/R$.  This leads to many interesting properties and
is an essential part of the web of dualities which relate
the different superstring theories and M theory.  One might ask whether it
is important that the dimension which is compactified is space-like.  
In this talk, I will
review some recent work\cite{Grignani:2000zm} which asks what happens
when the dimension that is compactified is light-like, rather than
space-like.

Of course, we could get a light-like circle by
boosting a compactified spatial circle by an infinite 
amount\cite{Seiberg:1997ad}. Consider a closed 
 string with compactified spatial
direction, $X^{D-1}\sim X^{D-1}+2\pi R$.  In terms of light-cone
coordinates, $X^\pm=\frac{1}{\sqrt{2}}\left( X^0\pm X^{D-1}\right)$
$$
\left(X^+,X^-\right)\sim \left( X^++\sqrt{2}\pi R,X^--\sqrt{2}\pi R
\right)
$$
We consider a boosted reference frame, where $\tilde X^+=\Lambda X^+$
and $\tilde X^- = \Lambda^{-1}X^-$.  We fix $\Lambda =\sqrt{2}R^+/R$
and take the limit $\Lambda\rightarrow\infty$ with $R^+$ fixed, we
finally get the light-like direction compactified, 
\begin{equation}
\left( \tilde
X^+,\tilde X^-\right)\sim \left( \tilde X^++{2}\pi R^+, X^-\right) 
\label{identification}
\end{equation}
The original spatial circle is vanishingly small, $R=
\sqrt{2}R^+/\Lambda$.
The momenta transform to
$$
P^+=\Lambda \frac{
\sqrt{(P^{D-1})^2+\vec P^2+M^2}+P^{D-1}}
{\sqrt{2}}$$
$$
P^-=\frac{
\sqrt{(P^{D-1})^2+\vec P^2+M^2}-P^{D-1} }
{\sqrt{2}\Lambda}$$
with $P^{D-1}=-N/R$ and $\vec p^2=\sum_1^{D-2}p_i^2$.  
In the limit of infinite boost this becomes
$$
\left( P^+,P^-\right)=
\left(   \frac{R^+}{2N}\left(\vec P^2+M^2\right) , N/R^+\right)
$$
Here, $P^+$ is the infinite momentum frame Hamiltonian which generates
translations of $X^-$.  Also, since $X^+\sim X^++2\pi R^+$ the conjugate
momentum is quantized, $P^+=N/R^+$.  Of the states with $N=0$, all but
the massless, low momentum ones get infinite energy.  

Closed string theory on a space with one dimension compactified to a
vanishingly small circle is T-dual to a closed string theory on the
un-compactified space.  As a consequence, the spatial compactification
to a vanishingly small circle 
has no effect on the spectrum of the theory.  For any light-like
compactification radius $R^+$, the energy spectrum of the rest-frame
states is just that of the de-compactified theory.

Of course the states which have finite energy and momentum in the
frame with compact light-cone must have infinite momentum in the
compact spatial direction and infinite energy in the original rest
frame.  Under T-duality, states with monzero momentum
in the vanishingly small compact 
direction are exchanged with states with fundamental strings wrapping
the very large dual circle.  Since they are very long, they have large
energy.  {\it Thus, light-like compactification does not alter the
spectrum of the string 
theory.  What it does is explores the theory which is T-dual to it in
a kinematical regime
where there are long fundamental strings wrapping an almost
infinite compact direction.}  This a high energy state in the rest frame
string theory and one could in principle study it there.  In the
infinite momentum frame it is a generic state with finite energy and
momentum.  This will be the reason why the zero temperature limit of the
partition functions that we shall compute in the following are independent 
of $R^+$.

The thermodynamic partition function is obtained from the trace over physical
states of the Boltzmann factor, $\exp (-\beta_\mu P^\mu)$
   \footnote{ Here,
we have introduced a covariant temperature.
$\beta_\mu$  should be understood
as the inverse temperature $\beta=1/k_B T=
\sqrt{2\beta_+\beta_--\vec\beta^2}$ times the D-velocity of
the heat bath $v^\mu=\beta^\mu/\beta$.
 $k_B$ is Boltzmann's constant.}
\begin{eqnarray}
Z=\sum_{N=0}^\infty e^{-\beta_- N/R^+}
\int \frac{d^{D-2}P}{(2\pi)^{D-2}}\cdot
e^{-\vec\beta\cdot\vec P} 
\nonumber\\
\cdot e^{-\beta_+ R^+\vec P^2/2N}
\sum_{M^2}\rho(M^2)e^{-\beta_+ R^+ M^2/2N}
\end{eqnarray}
where we sum over states in the mass spectrum.  These are conveniently
found in the light cone gauge by imposing the constraints $L_0 +
\tilde L_0 =0$ and the level matching condition $L_0-\tilde L_0
=N\cdot{\rm integer}$. (Details are explained in
\cite{Grignani:1999sp}.)  The first constraint gives the mass shell
condition.  The second is the level matching condition and is imposed
using an integer-valued Lagrange multiplier.  The result for the NSR
superstring is elegantly summarized in terms of the Hecke operator
\cite{serre} acting on the partition function of the superconformal
field theory with target space $R^8$,
\begin{equation}
-\frac{2\pi\beta_\mu R^\mu F}{V}={\cal
H}[e^{-\beta_\mu\beta^\mu/2\beta_\mu R^\mu
}]*{\cal F}[\tau,\bar\tau]
\label{p44}
\end{equation}
where
\begin{equation}
{\cal F}=\left[\left(\frac{1}{4\pi^2\alpha' \tau_2}\right)^4 
\frac{1}{\left| \eta(\tau)\right|^{24}}\left|
\theta_2(0,\tau)\right|^8\right]_{\tau=i\nu}
\label{pf4}
\end{equation}
and $ \nu=2\pi\alpha'/\beta_\mu R^\mu$ is a fixed constant ($\beta_\mu$
is spacelike and $R^\mu$ is light-like).
The factor in front contains 
the ratio of volumes of $R^8$ and $R^9\times
S^1$ with compactified light cone.  The action of ${\cal H}[p]$ on a
modular function $\phi(\tau,\bar\tau)$ is defined by
\begin{eqnarray}
{\cal H}[p]*\phi(\tau,\bar\tau)=
~~~~~~~~~~~~~~~~~~~~~~~~~~~
\nonumber \\
\sum_{N=1}^\infty p^N
\sum_{\stackrel {kr=N,~r~{\rm odd}} {s~{\rm mod}~k} }
\frac{1}{N}\phi\left(  \frac{s+\tau r}{k}, \frac{ s+\bar\tau  r}{k}\right)
\label{pf5}
\end{eqnarray}
This is similar to other partition functions for conformal field
theories on symmetric orbifolds. For a recent discussion see
ref.\cite{Fuji:2000fa}.

This result is a discretization of the usual Teichmuller space which
occurs in the genus 1 string amplitude.  Recall that the genus zero
contribution is insensitive to compactifications and for the
superstring it vanishes.  At genus 1, because of the modification of
the GSO projection by the finite temperature boundary conditions, the
superstring torus amplitude is non-zero.  Here, we see that the usual
integration over the Teichmuller space of tori is replaced by a summation
over discrete parameters
$$
\tau= \frac{s+i\nu r}{k}
$$

\section{A theorem about path integrals}

It is interesting to see what happens when we compactify
a null direction in the path integral representation of
the string free energy.  We can do this for the contribution
from arbitrary genus.  We begin with
\begin{equation}
F=-
\sum_{g=0}^\infty g_s^{2g-2} \int [dh_gdX]
e^{-S[h_g,X]}
\label{partf}
\end{equation}
where
\begin{equation}
S[h_g,X]=\frac{1}{4\pi\alpha'}\int \sqrt{h}h^{ab}\partial_aX^\mu\partial_b
X^\mu
\label{partf1}
\end{equation}
Here we will use the Bosonic string.  Most of our considerations apply
to the bosonic sector of any string theory.  For the relationship with
matrix theory which we shall discuss later, supersymmetry is important
and that case is more closely related to the Green-Schwarz
superstring.  The only modification of our arguments for the
superstring would be that either the world-sheet fermion boundary
conditions in the Green-Schwarz case, or the GSO projection in the
Neveu-Schwarz-Ramond case would be modified to depend on the winding
numbers of the world-sheet in the compact time direction.

The string
coupling constant is $g_s$ and its powers weight the genus,
of the string's world-sheet.  
For each value of the genus, $g$, $[dh_g]$ is an
integration measure over all metrics of that genus and is normalized
by dividing out the volume of the world-sheet re-parameterization and
Weyl groups.  We will assume that the metrics of both the world-sheet
and the target spacetime have Euclidean signatures.

We wish to study the situation where the target space has particular
compact dimensions.  Two compactifications will be needed.  The first
compactifies the light-cone in Minkowski space by
making the identification (\ref{identification}).  In our Euclidean
coordinates, 
\begin{equation}
\left( X^0, X^9\right)\sim \left( X^0+\sqrt{2}\pi iR, X^9 -\sqrt{2}
\pi R\right)
\label{comp1}
\end{equation}

In order to introduce temperature, we shall have to compactify Euclidean time,
\begin{equation}
\left( X^0,\vec X, X^9\right)\sim \left(X^0+\beta, \vec X, X^9\right)
\label{comp2}
\end{equation}
This compactification (with the appropriate modification of the GSO
projection in the case of superstrings) introduces the
temperature, $T=1/k_B\beta$, so that
(\ref{partf}) computes the thermodynamic free energy.

In order to implement this compactification in the path integral, we
assume that the world-sheet is a Riemann surface $\Sigma_g$ of genus
$g$ whose homology group $H_1(\Sigma_g)$ is generated by the
closed curves,
\begin{eqnarray}
a_1,a_2,\ldots, a_g~,~b_1, b_2, \ldots, b_g  \nonumber\\
a_i\cap a_j=\emptyset,~b_i\cap
b_j=\emptyset,~ a_i\cap b_j=\delta_{ij}
\label{homology}
\end{eqnarray}
Furthermore, one may pick a basis of holomorphic
differentials $\omega_i\in H^1(\Sigma_g)$ with the properties
\begin{equation}
\oint_{a_i}\omega_j=\delta_{ij}
~~~,~~~
\oint_{b_i}\omega_j=\Omega_{ij}
\label{orthog}
\end{equation}
where $\Omega$ is the period matrix.  It is complex, symmetric,
$\Omega_{ij}=\Omega_{ji}$, and has positive definite imaginary part.

Compactification is implemented by including the possible windings of
the string world-sheet on the compact dimensions.  These form distinct
topological sectors in the path integration in (\ref{partf}).  In the
winding sectors, the bosonic coordinates of the string should have a
multi-valued part which changes by $\beta\cdot$integer or
$(i)\sqrt{2}\pi R\cdot$integer as it is moved along a homology cycle.  The
derivatives of these coordinates should be single-valued functions.
It is convenient to consider their exterior derivatives which can be
expressed as linear combinations of the holomorphic and
anti-holomorphic 1-forms and exact parts,
\begin{eqnarray}
dX^0=\sum_{i=1}^{g}\left( \lambda_i\omega_i+\bar\lambda_i\bar\omega_i\right)+{\rm exact}
\nonumber \\
dX^9=\sum_{i=1}^{g}\left( \gamma_i\omega_i+\bar\gamma_i\bar\omega_i\right)+{\rm exact}
\label{coord}
\end{eqnarray}
Then, we require
\begin{eqnarray}
\oint_{a_i}dX^0=\beta n_i+\sqrt{2}\pi R^+ip_i
\nonumber \\
\oint_{b_i}dX^0=\beta m_i+\sqrt{2}\pi R^+i q_i
\nonumber\\
\oint_{a_i}dX^9=\sqrt{2}\pi R^+p_i
\nonumber \\
\oint_{b_i}dX^9=\sqrt{2}\pi R^+q_i
\end{eqnarray}
with $p_i,q_i,m_i,n_i$ integers.
With (\ref{orthog}), we use these equations to solve for the constants in
(\ref{coord}). With the formula
$
\int\omega_i \bar\omega_j
=-2i\left(\Omega_2\right)_{ij}
$, 
we compute the part of the string action which contains the winding integers,
\begin{eqnarray}
S=\frac{\beta^2}{4\pi\alpha'}\left(
n\Omega^{\dagger}-m\right)\Omega_2^{-1} \left( \Omega n-m\right)+ 
\nonumber \\
2\pi
i \frac{ \sqrt{2}\beta R^+}{4\pi\alpha'}\frac{1}{2}\left[\left(
p\Omega^{\dagger}-q\right)\Omega_2^{-1} \left( \Omega n-m\right)
\right. \nonumber \\ \left.  +\left(
n\Omega^{\dagger}-m\right)\Omega_2^{-1} \left( \Omega
p-q\right)\right]+\ldots
\end{eqnarray}
Note that the integers $p_i$ and $q_i$ appear linearly in a purely
imaginary term in the action. 
This is the only place that they
appear in the string path integral (unlike $m_i$ and $n_i$ which
could appear in the GSO projection).  When
the action is exponentiated and summed over $p_i$ and $q_i$, the
result will be periodic Dirac delta functions. 
These delta functions impose a linear constraint on the period matrix
of the world-sheet.  Thus, with the appropriate Jacobian factor, the
net effect is to insert into the path integral measure the following
expression,
\begin{eqnarray}
\sum_{\stackrel{mn}{rs}}e^{-\frac{\beta^2}{4\pi\alpha'}\left(
n\Omega^{\dagger}-m\right)\Omega_2^{-1} \left( \Omega n-m\right)}
\left| \det\Omega_2\right| 
\nonumber \\
\nu^{2g}
\prod_{j=1}^g
\delta\left(\left( n_i+i\nu r_i\right)\Omega_{ij}-\left(
m_j+i\nu s_j\right) \right)
\label{mod}
\end{eqnarray}
where $ \nu=4\pi\alpha'/\sqrt{2}\beta R^+$, the same constant as in 
(\ref{pf4}) if we specialize to the temperature D-vector $\beta_0\equiv
\beta$, all other components vanishing.
Consequently, the integration over metrics in the string
path integral is restricted to those for which the period matrix obeys
the constraint
\begin{equation}
\sum_{i=1}^g\left(n_i+i\nu r_i\right)\Omega_{ij}-\left(
m_j+i\nu s_j\right)=0
\label{const}
\end{equation}
for all combinations of the $4g$ integers $m_i,n_i,r_i,s_i$ such 
that $\Omega$ is in a fundamental domain of period matrices for surfaces
of genus $g$. 

Since the columns of the period matrix are linearly independent
vectors, these are $g$ independent complex constraints on the moduli
space of $\Sigma_g$.  Thus its complex dimension $3g-3$ is reduced to
$2g-3$ and there is further discrete data contained in the integers.
One would expect that, when the compactifications are removed, either
$\beta\rightarrow\infty$ or $R^+\rightarrow\infty$, the discrete data
assembles itself to a ``continuum limit'' which restores the 
complex dimension of moduli space. 

It is interesting to ask whether the Riemann surfaces with the
constraint (\ref{const}) can be classified in a sensible way.  The
answer to this question is yes, a Riemann surface obeys the constraint
(\ref{const}) if and only if it is a branched cover of the torus,
$T^2$, with Teichmuller parameter $i\nu$. This is established through the

\vspace{5pt}
\noindent
{\bf Theorem}: $\Sigma_{g}$ is a branched cover of $T^{2}$ if and
only if the period matrix obeys (\ref{const}), for some choice of integers
$m_{i}, n_{i}, r_{i}$ and $s_{i}$. 

\vspace{5pt}
\noindent
The {\it proof} can be found in ref. \cite{Grignani:2000zm}.

As a concrete example, the constraint (\ref{const}) can be solved
explicitly for genus one.  The torus amplitude for the finite temperature
type II superstring was given in the NSR formulation by Attick and Witten
\cite{aw}.

The modification of their formula by the null compactification can be
found using (\ref{mod}),
\begin{eqnarray}
\frac{F}{V}=-\sum_{\tau\in{\cal F}} 
\frac{\nu^2e^{-\frac{\beta^2 \vert n\tau-m\vert^2}{4\pi\alpha'\tau_2}}
}{m^2+\nu^2n^2} 
\left(\frac{1}{4\pi^2\alpha'\tau_2}\right)^5
\nonumber\\
\cdot \frac{1}{4\left| \eta(\tau)\right|^{24}}
\left[ \left( \theta_2^4\bar\theta_2^4
+\theta_3^4\bar\theta_3^4 + \theta_4^4\bar\theta_4^4\right)(0,\tau)
+
\right.
\nonumber\\
\left.
+e^{i\pi(m+n)}\left( \theta_2^4\bar\theta_4^4+\theta_4^4\bar\theta_2^4
\right)(0,\tau)
-e^{i\pi n}\left( \theta_2^4\bar\theta_3^4
\right.\right.
\nonumber\\
\left.\left.
+\theta_3^4\bar\theta_2^4
\right) (0,\tau)
- e^{i\pi m}\left( \theta_3^4\bar\theta_4^4 
+ \theta_4^4\bar\theta_3^4\right)(0,\tau)\right]\cr
\label{pf} 
\end{eqnarray}
where the solution of (\ref{const}) yields the discrete
Teichmuller parameter,
$$
\tau=\frac{m+i\nu s}{n+i\nu r}
$$
and one should sum over the integers so that $\tau$ is in the
fundamental domain, ${\cal F}$,
\begin{equation}{\cal F}\equiv\left\{\tau_1+i\tau_2 \left|
-\frac{1}{2}<\tau_1\leq\frac{1}{2};
\vert\tau\vert\geq 1; \tau_2>0 \right. \right\}
\label{fund}
\end{equation}

Modular transformations and identities for theta functions can be used
to rewrite (\ref{pf}) as the Hecke operator acting on the partition
function of a superconformal field theory, with torus world-sheet and
target space $R^8$ seen in the formulae (\ref{p44}), (\ref{pf4}) and
(\ref{pf5}).  

\section{Implications for the matrix model}

M-theory is a parameter free quantum mechanical system which has
11-dimensional super-Poincare symmetry, 
11-dimensional supergravity as its low energy limit and produces the
five known consistent superstring theories at various limits of its
moduli space.  Details of its dynamics are thus far unknown.  The
matrix model \cite{Banks:1997vh}, \cite{Susskind:1997cw} model is
conjectured to describe the full dynamics of M-theory in a particular
kinematical context, the infinite-momentum frame.

An important check of the matrix model conjecture would be to
use it to reproduce perturbative string theory.  Most straightforward
is the IIA superstring which is gotten by compactification of a
spatial direction of M-theory.  With this compactification, the matrix
model itself becomes 1+1-dimensional, maximally-supersymmetric
Yang-Mills theory.  According to Dijkgraaf, Verlinde and Verlinde
\cite{dvv}, the string degrees of freedom which emerge in the
perturbative string limit are simultaneous eigenvalues of the
matrices. At finite temperature, the matrices are defined on a torus
\cite{Grignani:1999sp} and their eigenvalues, since they solve polynomial
equations, are functions on branched covers of the torus.  If the
matrix model is to agree with perturbative string theory, these
branched covers must be the full set of Riemann surfaces that
contribute to the string path integral.  In the work that we have
reviewed here, we have indeed seen that this is the case, the moduli
spaces of branched covers which occur in the matrix model and the
world-sheets that occur in the measure in the string path integral are
identical.

\section*{Acknowledgments}
This work is supported in part by MaPhySto, Center for Mathematical Physics
and Stochastics, funded by the Danish National Research Foundation.  It is
also supported 
in part by NSERC of Canada.  I have benefited from conversations with
Jan Ambj\o rn, Gianluca Grignani, Mark Goresky, Janosh Kollar, Yutaka
Matsuo, Peter Orland, Lori Paniak, Savdeep Sethi, Graham Smith,
Richard Szabo, Erik Verlinde and Kostya Zarembo.


\begin{thebibliography}{99}

\bibitem{Grignani:2000zm}
G.~Grignani, P.~Orland, L.~D.~Paniak and G.~W.~Semenoff,
``Matrix theory interpretation of DLCQ string worldsheets,''
hep-th/0004194, Phys Rev. Lett. in press.

\bibitem{Seiberg:1997ad}
N.~Seiberg,
``Why is the matrix model correct?,''
Phys.\ Rev.\ Lett.\  {\bf 79}, 3577 (1997)
[hep-th/9710009].

\bibitem{Grignani:1999sp}
G.~Grignani and G.~W.~Semenoff,
``Thermodynamic partition function of matrix superstrings,''
Nucl.\ Phys.\  {\bf B561}, 243 (1999)
[hep-th/9903246].

\bibitem{serre}J.-P. Serre, {\it A Course in Arithmetic},
Springer-Verlag, New York, 1973.

\bibitem{Fuji:2000fa}
H.~Fuji and Y.~Matsuo,
``Open string on symmetric product,''
hep-th/0005111.

\bibitem{aw}J. Atick and E. Witten, ``The Hagedorn transition and the
number of degrees of freedom of string theory,'' Nucl.\ Phys.\ {\bf
B310}, 291 (1988).

\bibitem{Banks:1997vh}
T.~Banks, W.~Fischler, S.~H.~Shenker and L.~Susskind,
``M theory as a matrix model: A conjecture,''
Phys.\ Rev.\  {\bf D55}, 5112 (1997)
[hep-th/9610043].

\bibitem{Susskind:1997cw}
L.~Susskind,
``Another conjecture about M(atrix) theory,''
hep-th/9704080.

\bibitem{dvv}
R.~Dijkgraaf, E.~Verlinde and H.~Verlinde,
``Matrix string theory,''
Nucl.\ Phys.\  {\bf B500}, 43 (1997)
[hep-th/9703030].


\end{thebibliography}
\end{document}